\documentclass[preprint,preprintnumbers,nofootinbib]{revtex4-1}

\usepackage[normalem]{ulem}
\usepackage{graphicx}
\usepackage{dcolumn}
\usepackage{bm}
\usepackage{amsmath}
\usepackage{amsfonts}
\usepackage{latexsym}
\usepackage{bbm}
\usepackage{color}
\usepackage{amssymb}
\usepackage{amsthm}
\usepackage{epsfig}
\usepackage{physics}
\usepackage{hyperref}
\usepackage{comment}
\usepackage{ulem}
\usepackage{tikz-feynhand}
\hypersetup{
setpagesize=false,
 bookmarksnumbered=true,
 bookmarksopen=true,
 colorlinks=true,
 linkcolor=blue,
 citecolor=red,
}

\begin{document}

\preprint{\textbf{OCHA-PP-385}}

\title{Two Higgs doublet model with a complex singlet scalar and Multi-critical Point Principle}

\author{Gi-Chol Cho$^1$}
\email{cho.gichol@ocha.ac.jp}

\author{Chikako Idegawa$^2$}
\email{idegawa@mail.sysu.edu.cn}

\author{Chiaki Nose$^3$}
\email{chiaki.nose@hep.phys.ocha.ac.jp}

\affiliation{$^1$Department of Physics, Ochanomizu University, Tokyo 112-8610, Japan}

\affiliation{$^2$ MOE Key Laboratory of TianQin Mission,
TianQin Research Center for Gravitational Physics \& School of Physics and Astronomy,
Frontiers Science Center for TianQin,
Gravitational Wave Research Center of CNSA,
Sun Yat-sen University (Zhuhai Campus), Zhuhai 519082, China}

\affiliation{$^3$Graduate school of Humanities and Sciences, Ochanomizu University, Tokyo 112-8610, Japan}

\bigskip

\date{\today}

\begin{abstract}

We study a two Higgs doublet model extended by a complex singlet scalar, in which the imaginary part of the singlet serves as a dark matter (DM) candidate.
In this model, degenerate masses of the three neutral Higgs bosons are crucial for achieving consistency with current constraints from DM direct-detection experiments and Higgs searches. This is called the degenerate scalar scenario.
To provide a theoretical motivation for such a degenerate Higgs spectrum, we impose the tree-level Multiple Point Principle (MPP), which requires the electroweak and singlet vacua to be degenerate, and analyze its implications for the scalar potential, DM phenomenology, and the electroweak phase transition.
We show that the tree-level MPP favors large SU(2)$_L$ doublet-singlet mixing parameters, which compete with the degenerate scalar scenario.
Nevertheless, we demonstrate that viable parameter regions still exist in which the observed DM constraints are satisfied.
Furthermore, although the tree-level MPP forbids a tree-level-driven first-order electroweak phase transition, we show that thermal loop effects can induce a strong first-order transition compatible with electroweak baryogenesis.

\end{abstract}
\maketitle

%%%%%%%%%%%%%%%%%%%%%%%%%%%%%%%%%%%%%%%%%%%%%%%%
%		          Introduction
%%%%%%%%%%%%%%%%%%%%%%%%%%%%%%%%%%%%%%%%%%%%%%%%

\section{Introduction}\label{sec:intro}

The discovery of the Higgs boson in 2012 completed the Standard Model (SM) as a renormalizable quantum field theory describing electroweak symmetry breaking~\cite{ATLAS:2012yve,CMS:2012qbp}.
Nevertheless, the structure of the Higgs sector itself remains one of the most intriguing open questions in particle physics.
Among various extensions of the SM, the two Higgs doublet model (2HDM) is one of the simplest and most motivated frameworks~\cite{Lee:1973iz, Glashow:1976nt,Deshpande:1977rw, Branco:2011iw}.
However, the general form of the 2HDM does not provide a viable dark matter (DM) candidate, whereas the existence of DM has been firmly established by cosmological and astrophysical observations and cannot be explained within the SM.

To accommodate a DM candidate, we extend the 2HDM by introducing an additional complex singlet scalar field $S$~\cite{Jiang:2019soj,Zhang:2021alu,Biekotter:2021ovi,Biekotter:2022bxp, Dutta:2022qeq}.
We refer to this model as 2HDMS. In the 2HDMS, the imaginary part of the singlet behaves as a weakly interacting massive particle (WIMP) DM.
In addition, the neutral components of the two Higgs doublets and the real part of $S$ mix to form three Higgs bosons $H_{1,2,3}$.
Recent direct detection experiments, such as the LUX-ZEPLIN (LZ) experiment~\cite{LZ:2024zvo}, have placed stringent constraints on WIMP dark matter.
In this model, the elastic scattering between the DM particle and nucleons is mediated by three Higgs bosons.
Interestingly, the scattering cross section can be naturally suppressed when the Higgs masses are nearly degenerate~\cite{Cho:2024mea}.
This mechanism, known as the degenerate scalar scenario, provides a theoretically consistent explanation for the null results of direct detection experiments.

In previous studies, such mass degeneracy among scalars was introduced by hand.
In this work, we seek a principle that determines the scalar potential parameters and simultaneously realizes mass degeneracy.
For this purpose, we adopt the Multiple Point Principle (MPP)~\cite{Bennett:1993pj,Bennett:1996vy,Bennett:1996hx} as a guiding principle.
Originally, the MPP was proposed as a condition requiring the coexistence of multiple vacua with identical energy densities --- typically between the electroweak and a high-energy scale vacua --- and has successfully predicted the Higgs boson mass within the SM~\cite{Froggatt:1995rt}.
\begin{comment}
In the present model, however, the scalar potential naturally possesses two distinct vacua at the low-energy scale: one along the electroweak direction and the other along the singlet direction.
\end{comment}
In the present model, however, the scalar potential possesses two distinct vacua at the low-energy scale: one along the electroweak direction and the other along the singlet direction.
The authors of ref.~\cite{Kannike:2020qtw} argue that, if the MPP is a fundamental principle, it should be imposed on all possible vacua, including those dominated by the tree-level potential in models with extended scalar sectors.
We therefore consider the case in which these two vacua are degenerate and refer to this condition as the tree-level MPP. Such a setup not only provides an origin for the degenerate scalar scenario but also has important implications for the electroweak phase transition (EWPT)~\cite{Kuzmin:1985mm}. For a recent study, see ref.~\cite{Cho:2022zfg}.

We analyze the conditions imposed on the model parameters by the tree-level MPP and, at the same time, explore the parameter regions favored by the DM relic density and direct detection constraints.
Particular attention is paid to the mixing terms between the Higgs doublets and the singlet scalar, which play essential roles in both the MPP condition and the DM phenomenology, affecting the two in opposite directions.
We show that, even under these mutually competing constraints, there still exists an allowed parameter space where both conditions can be satisfied simultaneously.

On the other hand, once the tree-level MPP is imposed, the potential barrier relevant to the EWPT is not generated at tree level, and only one-loop thermal effects are responsible for its formation.
We demonstrate that, even under the tree-level MPP, a strong first-order EWPT required for successful electroweak baryogenesis can occur purely due to thermal loop effects.

The structure of this paper is as follows.
In Sec.~\ref{sec:CxSM}, we introduce the 2HDMS and define the scalar potential, mass spectrum, and relevant interactions.
In Sec.~\ref{sec:MPP}, we impose the tree-level MPP and discuss its implications for the vacuum structure and the EWPT.
In Sec.~\ref{sec:DM}, we study the DM phenomenology, focusing on the relic abundance and direct-detection constraints using representative benchmark points favored by the tree-level MPP.
Finally, Sec.~\ref{sec:sum} is devoted to a summary and conclusions.

%%%%%%%%%%%%%%%%%%%%%%%%%%%%%%%%%%%%%%%%%%%%%%%%
%		        Model
%%%%%%%%%%%%%%%%%%%%%%%%%%%%%%%%%%%%%%%%%%%%%%%%

\section{The Complex Singlet Extended Higgs model}\label{sec:CxSM}

\subsection{Model}\label{sec:model}

In the 2HDMS, the tree-level scalar potential $V_0$ is given by
\begin{align}
V_0 = V_{0,\mathrm{2HDM}} + V_{0,\mathrm{S}}.
\label{pot}
\end{align}
The first term on the right-hand side of Eq.~\eqref{pot} involves only the two Higgs doublet fields, $\Phi_1$ and $\Phi_2$:
\begin{align}
V_{0,\mathrm{2HDM}}\!\left(\Phi_1, \Phi_2\right)
= &\, m_1^2 \Phi_1^{\dagger} \Phi_1 + m_2^2 \Phi_2^{\dagger} \Phi_2
- \left( m_3^2 \Phi_1^{\dagger} \Phi_2 + \text{h.c.} \right) \nonumber \\
& + \frac{\lambda_1}{2} (\Phi_1^{\dagger} \Phi_1)^2
+ \frac{\lambda_2}{2} (\Phi_2^{\dagger} \Phi_2)^2
+ \lambda_3 (\Phi_1^{\dagger} \Phi_1)(\Phi_2^{\dagger} \Phi_2) \nonumber \\
& + \lambda_4 (\Phi_1^{\dagger} \Phi_2)(\Phi_2^{\dagger} \Phi_1)
+ \left[ \frac{\lambda_5}{2} (\Phi_1^{\dagger} \Phi_2)^2 + \text{h.c.} \right],
\end{align}
where only the $m_3^2$ term softly breaks the $Z_2$ symmetry of the doublets
($\Phi_1 \to +\Phi_1$, $\Phi_2 \to -\Phi_2$).
The second term in Eq.~\eqref{pot}, involving the singlet field $S$ and the doublets, is given by
\begin{align}
V_{0,\mathrm{S}}\!\left(\Phi_1, \Phi_2, S\right)
&= \frac{\delta_1}{2} \Phi_1^{\dagger} \Phi_1 |S|^2
+ \frac{\delta_2}{2} \Phi_2^{\dagger} \Phi_2 |S|^2
+ \frac{b_2}{2} |S|^2
+ \frac{d_2}{4} |S|^4 \nonumber \\
&\quad + \left( a_1 S + \frac{b_1}{4} S^2 + \text{h.c.} \right),
\end{align}
where the terms in the first line are invariant under a global U(1) transformation
$S \to e^{i\theta} S$, while those in the second line softly break this symmetry.
In the following, all coefficients in the scalar potential are taken to be real.

The Higgs doublets $\Phi_i~(i=1,2)$ and the singlet $S$ are expanded around their vacuum expectation values (VEVs) as
\begin{align}
\Phi_i =
\begin{pmatrix}
\phi_i^{+} \\
\frac{1}{\sqrt{2}}\,(v_i + h_i + i \eta_i)
\end{pmatrix},
\qquad
S = \frac{1}{\sqrt{2}}\,(v_S + s + i \chi).
\end{align}
Here $v \equiv \sqrt{v_1^2 + v_2^2} = 246.22~\mathrm{GeV}$ and $\tan\beta \equiv v_2 / v_1$.
The components of $\Phi_i$ and $S$ consist of charged scalars $(\phi_i^{+})$, neutral scalar components $(h_1, h_2, s)$, and neutral pseudoscalar components $(\eta_i, \chi)$.
The field $\chi$ is stable and serves as the DM candidate.
The first derivatives of $V_0$ with respect to $h_1$, $h_2$, and $s$ yield the following tadpole conditions:
\begin{align}
\left\langle \frac{\partial V_0}{\partial h_1} \right\rangle
&= m_1^2 v_1 - m_3^2 v_2
+ \frac{\lambda_1}{2} v_1^3
+ \frac{\lambda_{345}}{2} v_1 v_2^2
+ \frac{\delta_1}{4} v_1 v_S^2 = 0, \label{tad1} \\
\left\langle \frac{\partial V_0}{\partial h_2} \right\rangle
&= m_2^2 v_2 - m_3^2 v_1
+ \frac{\lambda_2}{2} v_2^3
+ \frac{\lambda_{345}}{2} v_1^2 v_2
+ \frac{\delta_2}{4} v_2 v_S^2 = 0, \label{tad2} \\
\left\langle \frac{\partial V_0}{\partial s} \right\rangle
&= \sqrt{2}\,a_1 + \frac{b_1 + b_2}{2} v_S
+ \frac{\delta_1}{4} v_1^2 v_S
+ \frac{\delta_2}{4} v_2^2 v_S
+ \frac{d_2}{4} v_S^3 = 0, \label{tad3}
\end{align}
where $\lambda_{345} = \lambda_3 + \lambda_4 + \lambda_5$.

The mass matrix for the neutral scalar components is obtained as
\begin{align}
-\mathcal{L}_{\text{mass}}
&= \frac{1}{2}
\begin{pmatrix}
h_1 & h_2 & s
\end{pmatrix}
\mathcal{M}_{S}^{2}
\begin{pmatrix}
h_1 \\ h_2 \\ s
\end{pmatrix}
= \frac{1}{2}
\begin{pmatrix}
H_1 & H_2 & H_3
\end{pmatrix}
O^{\top} \mathcal{M}_{S}^{2} O
\begin{pmatrix}
H_1 \\ H_2 \\ H_3
\end{pmatrix}
= \frac{1}{2} \sum_{i=1}^3 m_{H_i}^2 H_i^2,
\label{mixingrelation}
\end{align}
where
\begin{align}
\mathcal{M}_S^2
= \begin{pmatrix}
m_3^2 \frac{v_2}{v_1} + \lambda_1 v_1^2 & -m_3^2 + \lambda_{345} v_1 v_2 & \frac{\delta_1}{2} v_1 v_S \\
 -m_3^2 + \lambda_{345} v_1 v_2 & m_3^2 \frac{v_1}{v_2} + \lambda_2 v_2^2 & \frac{\delta_2}{2} v_2 v_S \\
 \frac{\delta_1}{2} v_1 v_S & \frac{\delta_2}{2} v_2 v_S & -\frac{\sqrt{2} a_1}{v_S} + \frac{d_2}{2} v_S^2
\end{pmatrix}.
\end{align}
The orthogonal matrix $O$ diagonalizes $\mathcal{M}_S^2$ and can be parametrized as
\begin{align}
O(\alpha_i)
&=
\begin{pmatrix}
1 & 0 & 0 \\[3pt]
0 & c_3 & -s_3 \\[3pt]
0 & s_3 & c_3
\end{pmatrix}
\begin{pmatrix}
c_2 & 0 & -s_2 \\[3pt]
0 & 1 & 0 \\[3pt]
s_2 & 0 & c_2
\end{pmatrix}
\begin{pmatrix}
c_1 & -s_1 & 0 \\[3pt]
s_1 & c_1 & 0 \\[3pt]
0 & 0 & 1
\end{pmatrix},
\end{align}
where $s_i = \sin\alpha_i$ and $c_i = \cos\alpha_i$ $(i = 1,2,3)$.
The orthogonality condition is given by
\begin{align}
\sum_k O_{ik} O_{jk} = \delta_{ij}.
\label{orthgnl}
\end{align}
The three mass eigenstates $H_{1,2,3}$ correspond to the physical neutral Higgs bosons, where $H_1$ is identified with the observed 125~GeV Higgs at the LHC.
On the other hand, the charged scalar $H^{\pm}$ and the pseudoscalar $A$ are defined from $\phi_{1,2}^{+}$ and $\eta_{1,2}$ as
\begin{align}
\begin{pmatrix}
\phi_1^{+} \\ \phi_2^{+}
\end{pmatrix}
= R(\beta)
\begin{pmatrix}
G^{+} \\ H^{+}
\end{pmatrix},
\qquad
\begin{pmatrix}
\eta_1 \\ \eta_2
\end{pmatrix}
= R(\beta)
\begin{pmatrix}
G^0 \\ A
\end{pmatrix},
\label{mssgn}
\end{align}
with
\begin{align}
R(\beta)
= \begin{pmatrix}
\cos\beta & -\sin\beta \\
\sin\beta & \cos\beta
\end{pmatrix}.
\end{align}
Here $G^{\pm}$ and $G^0$ denote the Nambu-Goldstone bosons.
The corresponding mass eigenvalues are given by
\begin{align}
m_{H^{\pm}}^2 &= \frac{m_3^2}{\sin\beta \cos\beta}
- \frac{1}{2}(\lambda_4 + \lambda_5) v^2,
\label{mHch} \\
m_A^2 &= \frac{m_3^2}{\sin\beta \cos\beta}
- \lambda_5 v^2,
\label{mA}
\end{align}
and the DM mass is
\begin{align}
m_{\chi}^2 = -\frac{\sqrt{2} a_1}{v_S} - b_1.
\label{mDM}
\end{align}

Let us summarize the input parameters and several theoretical constraints.
The scalar potential contains 14 parameters $\{m_1^2, m_2^2, m_3^2, \lambda_1, \lambda_2, \lambda_3, \lambda_4, \lambda_5,
\delta_1, \delta_2, b_2, d_2, a_1, b_1\}.$
Among them, $\{m_3^2, a_1\}$ are treated as free input parameters, while the remaining 12 parameters are output parameters.
The parameters $\{m_1^2, m_2^2, b_2\}$ are determined by the tadpole
conditions~\eqref{tad1}--\eqref{tad3}, and
$\{\lambda_4, \lambda_5, b_1\}$ are fixed by the physical masses in
Eqs.~\eqref{mHch}--\eqref{mDM}.
The remaining six parameters are determined from the neutral scalar mass matrix.
%----------------------------------------------------------------------------------------------------------------------------------
\begin{table}[t]
\centering
\begin{tabular}{c|cccccccccccccc}
\hline\hline
Inputs & $v,$ & $v_S,$ & $m_{H_1},$ & $m_{H_2},$ & $m_{H_3},$ & $m_{H^\pm},$ & $m_{A},$ & $m_{\chi},$ & $\alpha_1,$ & $\alpha_2,$ & $\alpha_3,$ & $\tan\beta,$ & $m_3^2,$ & $a_1$\\
\hline
Outputs & $v_1,$ & $v_2,$ & $m_1^2,$ & $m_2^2,$ & $\lambda_1,$ & $\lambda_2,$ & $\lambda_3,$ & $\lambda_4,$ & $\lambda_5,$ & $\delta_1,$ & $\delta_2,$ & $b_2,$ & $d_2,$ & $b_1$ \\
\hline\hline
\end{tabular}
\caption{Input and output parameters in this model.}
\label{tab:inout}
\end{table}
%----------------------------------------------------------------------------------------------------------------------------------
The input and output parameters are summarized in Table~\ref{tab:inout}.

Since $\delta_1$, $\delta_2$, and $d_2$ play a central role in the discussion of the phase transition, we display their explicit expressions here:
\begin{align}
\delta_1 &= \frac{2}{v_1 v_S} \sum_{i=1}^3 O_{1i} O_{3i} m_{H_i}^2,\label{del1}
\\
\delta_2 &= \frac{2}{v_2 v_S} \sum_{i=1}^3 O_{2i} O_{3i} m_{H_i}^2,\label{del2}
\\
d_2 &= \frac{2}{v_S^2}
\left(
    \frac{\sqrt{2}\, a_1}{v_S}
    + \sum_{i=1}^3 O_{3i}^2 m_{H_i}^2
\right).\label{d2}
\end{align}
These relations will be relevant when discussing both the MPP condition and the structure of the EWPT.
All remaining relations between the physical inputs and the original Lagrangian
parameters are summarized in Appendix~\ref{app:para}.

The scalar potential is required to be bounded from below, leading to the conditions
\cite{Nie:1998yn,Kanemura:1999xf}
\begin{align}
\lambda_1 > 0,\qquad
\lambda_2 > 0,\qquad
\lambda_3 + \lambda_4 - \lambda_5 > -\sqrt{\lambda_1 \lambda_2},\qquad
d_2 > 0.
\end{align}
The quartic couplings are further constrained by tree-level unitarity
\cite{Kanemura:1993hm, Akeroyd:2000wc, Ginzburg:2005dt, Chen:2014ask, Maleki:2022zuw}:
\begin{align}
\lambda_1 < \frac{8\pi}{3},\qquad
\lambda_2 < \frac{8\pi}{3},\qquad
\lambda_{345} < 8\pi,\qquad
\delta_1 < 16\pi,\qquad
\delta_2 < 16\pi,\qquad
d_2 < \frac{16\pi}{3}.
\end{align}
Perturbativity additionally requires \cite{Aoki:2021oez}
\begin{align}
|\lambda_i|,\, d_2 < 4\pi \qquad (i = 1\text{--}5).
\end{align}

%----------------------------------------------------------------------------------------------------------------------------------
\begin{table}[t]
\centering
\begin{tabular}{|c|c|c|c|c||c|c|c|}
\hline
 & $Q_L, L_L$ & $u_R$ & $d_R$ & $\ell_R$ & $\Phi_u$ & $\Phi_d$ & $\Phi_\ell$ \\ \hline
Type-I  & $+$ & $-$ & $-$ & $-$ & $\Phi_2$ & $\Phi_2$ & $\Phi_2$ \\ \hline
Type-II & $+$ & $-$ & $+$ & $+$ & $\Phi_2$ & $\Phi_1$ & $\Phi_1$ \\ \hline
Type-X  & $+$ & $-$ & $-$ & $+$ & $\Phi_2$ & $\Phi_2$ & $\Phi_1$ \\ \hline
Type-Y  & $+$ & $-$ & $+$ & $-$ & $\Phi_2$ & $\Phi_1$ & $\Phi_2$ \\ \hline
\end{tabular}
\caption{
Assignments of $Z_2$ charges to the fermions and the corresponding Higgs doublets that couple to each fermion type.
The Higgs doublets transform as $\Phi_1 \to +\Phi_1$ and $\Phi_2 \to -\Phi_2$, respectively.
}
\label{tab:Z2charge}
\end{table}
%----------------------------------------------------------------------------------------------------------------------------------

To avoid tree-level flavor-changing neutral currents (FCNCs), we assume that each type of fermion couples to only one of the Higgs doublets~\cite{Glashow:1976nt,Paschos:1976ay}.
The Yukawa Lagrangian is written as
\begin{align}
-\mathcal{L}_{\text{Yukawa}}
= \bar{Q}_L Y_u \tilde{\Phi}_u u_R
+ \bar{Q}_L Y_d \Phi_d d_R
+ \bar{L}_L Y_{\ell} \Phi_{\ell} \ell_R
+ \text{h.c.},
\end{align}
where $\Phi_f~(f = u, d, \ell)$ is the Higgs doublet that couples to fermion $f$,
$\tilde{\Phi}_u \equiv i\sigma_2 \Phi_u^*$, and $\sigma_2$ is the Pauli matrix.
Here $Q_L = (u_L, d_L)^{\top}$ and $L_L = (\nu_L, e_L)^{\top}$ represent the left-handed quark and lepton doublets, respectively, while $u_R$, $d_R$, and $\ell_R$ are the corresponding right-handed singlets.
$Y_f~(f = u, d, \ell)$ denote the $3\times3$ Yukawa coupling matrices.
As summarized in Table~\ref{tab:Z2charge}, by assigning appropriate $Z_2$ charges to the fermions, one can identify which Higgs doublet couples to each fermion type, leading to four distinct Yukawa structures.

\subsection{Degenerate scalar scenario}\label{subsec:DSS}

Recent DM direct detection experiments have provided stringent upper limits
on the spin-independent DM--nucleon scattering cross section~\cite{LZ:2024zvo},
which strongly constrain various Higgs-portal-type DM models.
In the 2HDMS, the elastic scattering between the DM particle $\chi$ and nucleons is mediated by the three neutral Higgs bosons $H_i$ $(i=1,2,3)$.
The relevant process is
\begin{align}
 \chi(p_1) + q(p_2) \rightarrow \chi(p_3) + q(p_4),
\end{align}
which proceeds through the $t$-channel exchange of $H_1$, $H_2$, and $H_3$.

As an illustrative example, we consider the Type-I 2HDMS, in which all fermions couple only to the Higgs doublet $\Phi_2$.
The couplings between the up-type and the down-type quark $u, d$ and the mass eigenstates $H_i$ are therefore written as
\begin{align}
-\mathcal{L}_{\rm Yukawa}
=
\sum_{i=1}^{3} C_{uuH_i}\, \overline{u_L}u_R H_i
+\sum_{i=1}^{3} C_{ddH_i}\, \overline{d_L}d_R H_i
+\text{h.c.},
\end{align}
with
\begin{align}
C_{uuH_i}=\frac{m_u}{v_2} O_{2 i},\qquad
C_{ddH_i}=\frac{m_d}{v_2} O_{2 i},
\end{align}
where $m_{u,d}$ denote the up- and down-type quark masses.
Since we focus on the DM--quark scattering, the lepton sector is omitted.
The trilinear scalar interactions involving the DM field $\chi$ are written as
\begin{align}
-\mathcal{L}\supset
C_{\chi\chi H_i}\, H_i \chi^2,
\end{align}
where
\begin{align}
C_{\chi\chi H_i}
= C_{\chi\chi h_1} O_{1i}
+ C_{\chi\chi h_2} O_{2i}
+ C_{\chi\chi s} O_{3i}.
\label{ccssdd_new}
\end{align}
Explicit expressions of these couplings are given by
\begin{align}
C_{\chi\chi h_1}&=\frac{\delta_1}{4}v_1,\quad C_{\chi\chi h_2}=\frac{\delta_2}{4}v_2,\quad C_{\chi\chi s}=\frac{d_2}{4}v_S.
\end{align}

In direct detection experiments, the squared momentum transfer $t \equiv (p_1-p_3)^2$ is much smaller than the mediator masses, $t\ll m_{H_i}^2$.
Under this approximation, the amplitudes for up- and down-type quarks become
\begin{align}
i\mathcal{M}_{\rm up}
&=
2 i\, \bar{u}(p_4)u(p_2)\,
\frac{m_u}{v_2}
\sum_{i=1}^{3}
\frac{C_{\chi\chi H_i} O_{2i}}{m_{H_i}^2}
\nonumber \\[3pt]
&=
2 i\, \bar{u}(p_4)u(p_2)\,
\frac{m_u}{v_2}
\sum_{i=1}^{3}
\left(
C_{\chi\chi h_1}\frac{O_{1i}O_{2i}}{m_{H_i}^2}
+ C_{\chi\chi h_2}\frac{O_{2i}^2}{m_{H_i}^2}
+ C_{\chi\chi s}\frac{O_{3i}O_{2i}}{m_{H_i}^2}
\right),
\label{cndtn1_new}
\\[6pt]
i\mathcal{M}_{\rm down}
&=
2 i\, \bar{u}(p_4)u(p_2)\,
\frac{m_d}{v_2}
\sum_{i=1}^{3}
\left(
C_{\chi\chi h_1}\frac{O_{1i}O_{2i}}{m_{H_i}^2}
+ C_{\chi\chi h_2}\frac{O_{2i}^2}{m_{H_i}^2}
+ C_{\chi\chi s}\frac{O_{3i}O_{2i}}{m_{H_i}^2}
\right).
\label{cndtn2_new}
\end{align}

If the three Higgs boson masses are degenerate, the orthogonality of the mixing matrix implies
\[
\sum_i O_{1i}O_{2i}=0,
\qquad
\sum_i O_{3i}O_{2i}=0,
\]
so the terms proportional to $O_{1i}O_{2i}$ and $O_{3i}O_{2i}$ cancel exactly.
The only surviving contribution is the one proportional to $O_{2i}^2$, whose size is controlled by
\begin{align}
C_{\chi\chi h_2}=\frac{\delta_2}{4}\, v_2.
\end{align}
The coefficient $\delta_2$ is given by Eq.~\eqref{del2}, and is suppressed because
the sum $\sum_i O_{2i} O_{3i} m_{H_i}^2$ becomes small due to the orthogonality of the mixing matrix.
Hence, in the degenerate scalar limit, the entire amplitude becomes strongly suppressed.
This orthogonality-induced suppression of $\delta_2$ is the essential ingredient of the degenerate scalar scenario.
However, if either $v_2$ or $v_S$ is too small, $\delta_2$ cannot be reduced, and the mechanism may cease to be effective.
In Type-X Yukawa interactions, the same $\delta_2$-suppression pattern persists.
In Type-II and Type-Y, on the other hand, the up-type (down-type) quarks couple with $\Phi_2$ ($\Phi_1$). Following a similar discussion, both Type-II and Type-Y require not only $\delta_2$-suppression, but also $\delta_1$-suppression. For a more detailed discussion, see ref.~\cite{Cho:2024mea}.

We also comment on an important phenomenological aspect of the degenerate scalar scenario relevant for Higgs searches at the LHC.
In the 2HDMS, the couplings of the neutral Higgs bosons $H_i$ to the SM final state $X$ are given by the corresponding SM Higgs coupling multiplied by the mixing factor $O_{1i}$.
The partial decay widths of $H_i$ can be written as
\begin{align}
\Gamma(H_1 \to XX) &= \Gamma_{h \to XX}^{\mathrm{SM}}(m_{H_1})\, O_{11}^2, \\
\Gamma(H_2 \to XX) &= \Gamma_{h \to XX}^{\mathrm{SM}}(m_{H_2})\, O_{12}^2, \\
\Gamma(H_3 \to XX) &= \Gamma_{h \to XX}^{\mathrm{SM}}(m_{H_3})\, O_{13}^2,
\end{align}
where $\Gamma_{h\to XX}^{\mathrm{SM}}(m_{H_i})$ denotes the SM Higgs decay width evaluated at $m_{H_i}$.
If the three Higgs bosons are nearly degenerate around the observed Higgs mass, $m_{H_i}\simeq m_h = 125~\mathrm{GeV}$, their combined decay width becomes
\begin{align}
\Gamma(H_1 \to XX) + \Gamma(H_2 \to XX) + \Gamma(H_3 \to XX)
\simeq \Gamma_{h\to XX}^{\mathrm{SM}}(m_h),
\end{align}
because of the orthogonality relation $\sum_{i=1}^3 O_{1i}^2 = 1$.
Consequently, the inclusive Higgs signal rate remains indistinguishable from the SM prediction for any values of the mixing angles, and the individual states $H_i$ cannot be resolved with the current LHC sensitivity.

%%%%%%%%%%%%%%%%%%%%%%%%%%%%%%%%%%%%%%%%%%%%%%%%
%	        Tree-level MPP and EWPT
%%%%%%%%%%%%%%%%%%%%%%%%%%%%%%%%%%%%%%%%%%%%%%%%

\section{Tree-level MPP and EWPT}\label{sec:MPP}

In this section, we discuss the application of the MPP to the 2HDMS at the tree level and examine its implications for the EWPT.
The MPP requires the coexistence of multiple vacua with identical energy densities.
While the original idea of the MPP was formulated between the electroweak and a high-energy vacuum in the SM~\cite{Bennett:1993pj,Bennett:1996vy,Bennett:1996hx}, in the present model,f the scalar potential naturally possesses two distinct vacua at the low-energy scale: one along the electroweak direction and the other along the singlet direction.
We therefore impose the tree-level MPP condition, which requires these two vacua to be degenerate in energy, as proposed in Ref.~\cite{Kannike:2020qtw}.
This condition serves as a basis for examining possible implications for the degenerate scalar scenario and the EWPT.

The classical background fields of the Higgs doublets and the singlet are parameterized as
\begin{align}
\langle \Phi_i \rangle &=
\frac{1}{\sqrt{2}}
\begin{pmatrix}
0 \\ \varphi_i
\end{pmatrix},
\qquad
\langle S \rangle = \frac{\varphi_S}{\sqrt{2}}.
\end{align}
The tree-level potential in Eq.~\eqref{pot} can then be expressed in terms of these background fields as
\begin{align}
V_0(\varphi_1, \varphi_2, \varphi_S)
&= \frac{m_1^2}{2}\,\varphi_1^2
+ \frac{m_2^2}{2}\,\varphi_2^2
- m_3^2\,\varphi_1 \varphi_2
+ \frac{\lambda_1}{8}\,\varphi_1^4
+ \frac{\lambda_2}{8}\,\varphi_2^4
+ \frac{\lambda_{345}}{4}\,(\varphi_1 \varphi_2)^2 \nonumber \\
&\quad
+ \frac{\delta_1}{8}\,\varphi_1^2 \varphi_S^2
+ \frac{\delta_2}{8}\,\varphi_2^2 \varphi_S^2
+ \frac{b_2}{4}\,\varphi_S^2
+ \frac{d_2}{16}\,\varphi_S^4
+ \sqrt{2}\,a_1\,\varphi_S
+ \frac{b_1}{4}\,\varphi_S^2.
\end{align}
The pattern of the EWPT in the 2HDMS corresponds to the transition from the singlet vacuum $(0,0,v_S')$ to the electroweak vacuum $(v_1, v_2, v_S)$.
This structure originates from the scalar potential, in which the stationary conditions for the singlet direction differ from those for the electroweak direction due to the absence of the $\delta_{1,2}$ terms in the singlet vacuum, together with the presence of the linear term $a_1 S$, which shifts the singlet direction away from the origin. 
As a result, the potential generically admits two distinct extrema corresponding to the electroweak and singlet vacua, without requiring special parameter choices (see also Eq.~\eqref{tadpoles}). 
These features also have important implications for the thermal history. In particular, due to the linear term $a_1 S$, the singlet direction is generically shifted away from the origin and is not driven to the symmetric point even at high temperatures. This makes it plausible that the singlet vacuum is realized prior to the electroweak vacuum.
To discuss the EWPT, we work in the Landau gauge and assume that the charged scalar fields do not acquire VEVs at any temperature, so that the U(1)$_\mathrm{QED}$ symmetry remains conserved.

The difference between the energy densities at the two vacua is expressed as
\begin{align}
\Delta V_0 &\equiv V_0(v_1, v_2, v_S) - V_0(0, 0, v_S')
\nonumber \\
&= \frac{m_1^2}{4} v_1^2
+ \frac{m_2^2}{4} v_2^2
- \frac{m_3^2}{2} v_1 v_2
+ \frac{3 \sqrt{2}}{4} a_1 (v_S - v_S^{\prime})
+ \frac{1}{8}(b_1 + b_2)(v_S^2 - v_S^{\prime 2}),
\label{deltaV}
\end{align}
where the tree-level MPP requires $\Delta V_0 = 0$.
Electroweak symmetry breaking occurs when the quadratic coefficients $m_1^2$ and $m_2^2$ take negative values,
leading to nonzero vacuum expectation values $v_{1,2}$.
In our analysis, $m_3^2$ is treated as an input parameter and assumed to be positive.
Under these assumptions, the first three terms in Eq.~\eqref{deltaV} are negative.
To realize the tree-level MPP, the fourth and fifth terms in Eq.~\eqref{deltaV} must compensate  for this deficit.
Since they scale with $(v_S - v_S')$ and $(v_S^2 - v_S'^2) = (v_S - v_S')(v_S + v_S')$,
a larger separation $|v_S - v_S'|$ tends to enhance their effect;
the net sign, however, depends on $a_1$ and $b_1 + b_2$ (and on $v_S + v_S'$).

Two VEVs, $v_S$ and $v_S'$, are obtained from the stationary conditions of the potential as
\begin{align}
\frac{d_2}{2} v_S^3
+ \left(b_1 + b_2 + \frac{\delta_1}{2} v_1^2 + \frac{\delta_2}{2} v_2^2 \right) v_S
+ 2 \sqrt{2}\, a_1 &= 0, \nonumber \\
\frac{d_2}{2} v_S^{\prime 3}
+ \left(b_1 + b_2 \right) v_S^{\prime}
+ 2 \sqrt{2}\, a_1 &= 0.
\label{tadpoles}
\end{align}
In the singlet vacuum, the Higgs doublets do not acquire VEVs, and therefore the terms proportional to $\delta_{1,2}$ are absent in the stationary condition.
This difference between the two equations determines the distinct values of $v_S$ and $v_S'$.
Consequently, increasing $\delta_{1,2}$ tends to enlarge the separation between $v_S$ and $v_S'$, which in turn facilitates the realization of $\Delta V_0 = 0$ under the tree-level MPP condition.

The mixing parameters $\delta_{1,2}$ are given in terms of the mass eigenvalues
and the mixing matrix by Eqs.~\eqref{del1} and \eqref{del2}.
In the degenerate scalar scenario, the combinations
$\sum_{i=1}^3 O_{1i} O_{3i} m_{H_i}^2$ and
$\sum_{i=1}^3 O_{2i} O_{3i} m_{H_i}^2$ become suppressed due to the
orthogonality of the mixing matrix and the near degeneracy among the scalar masses.
Therefore, obtaining large values of $\delta_{1,2}$ requires a relatively small
singlet VEV $v_S$, since $v_1$ and $v_2$ are fixed by $v_1^2 + v_2^2 = v^2$
and cannot be reduced.

%----------------------------------------------------------------------------------------------------------------------------------
\begin{table}[t]
\centering
\begin{tabular}{|c|c|c|c|c|c|c|c|c|c|c|c|}
\hline
 $v$ & $m_{H_1}$ & $m_{H_2}$ & $m_{H_3}$ & $m_{H^\pm}$ & $m_{A}$ & $m_{\chi}$ & $\alpha_1$ & $\alpha_2$ & $\alpha_3$ & $\tan\beta$ & $m_3^2$ \\
\hline
 246.22 & 125.0 & 124.5 & 124.0 & 500 & 500 & 200 & $\pi/4$ & $\pi/4$ & 0.01 & 2.0 & 10 \\
\hline
\end{tabular}
\caption{Input parameters adopted for the degenerate scalar scenario.
The singlet VEV $v_S$ and the linear parameter $a_1$ are treated as free parameters and scanned in the numerical analysis. The units of VEV and masses are GeV, while that of $m_3^2$ is GeV$^2$. }
\label{tab:input}
\end{table}
%----------------------------------------------------------------------------------------------------------------------------------

Before presenting the numerical results for the tree-level MPP condition,
we show the parameter setup adopted for the degenerate scalar scenario.
Representative values consistent with the degenerate scalar scenario and phenomenological constraints are summarized in Table~\ref{tab:input}.
The masses of the three Higgs bosons, $m_{H_i}$, are degenerate within 1~GeV.
As discussed in Sec.~\ref{subsec:DSS}, when the Higgs masses are nearly degenerate,
the mixing angles $\alpha_i$ are not strongly constrained.

In the following analysis, we adopt the Type-I Yukawa structure,
in which all fermions couple to the same Higgs doublet $\Phi_2$.
Experimental constraints are primarily associated with electroweak precision data and
flavor observables.
By assuming a mass degeneracy between the charged scalar $H^\pm$ and either the neutral
scalar $H$ or the pseudoscalar $A$, i.e., $m_{H^\pm} \simeq m_H$ or $m_A$,
the electroweak precision data can be satisfied~\cite{Haber:2010bw}.
In the Type-I 2HDMS, the process $B_d \rightarrow \mu^+ \mu^-$ requires
$\tan\beta \gtrsim 1.75$ for $m_{H^\pm} = 500~\mathrm{GeV}$~\cite{Haller:2018nnx}.
We note that Higgs coupling measurements usually constrain $\cos(\beta-\alpha)$; however,
this constraint does not apply in our analysis because it is based on the degenerate
scalar scenario.

%----------------------------------------------------------------------------------------------------------------------------------
\begin{figure}[htpb]
\centering
\includegraphics[width=8cm]{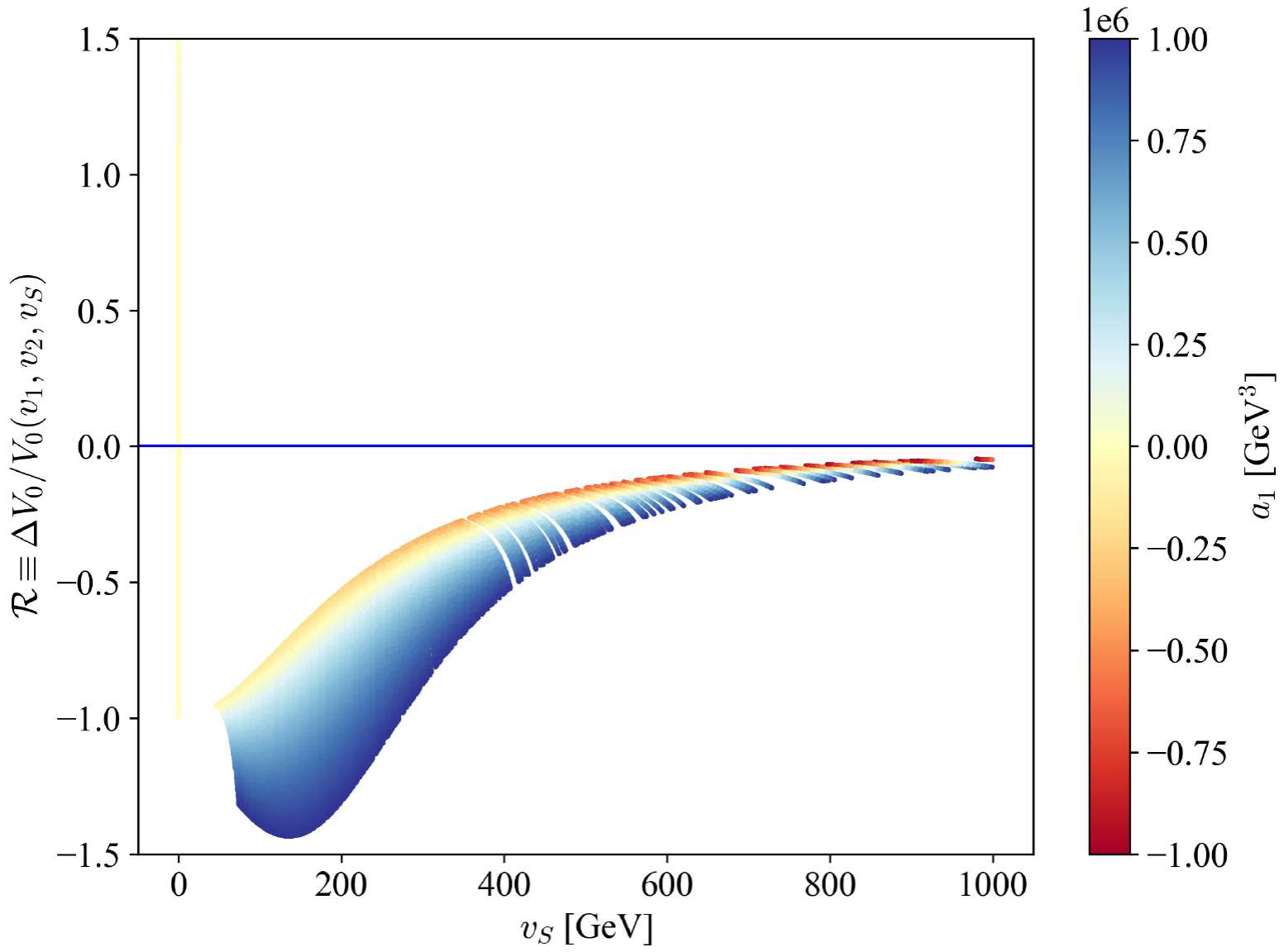}
\includegraphics[width=8cm]{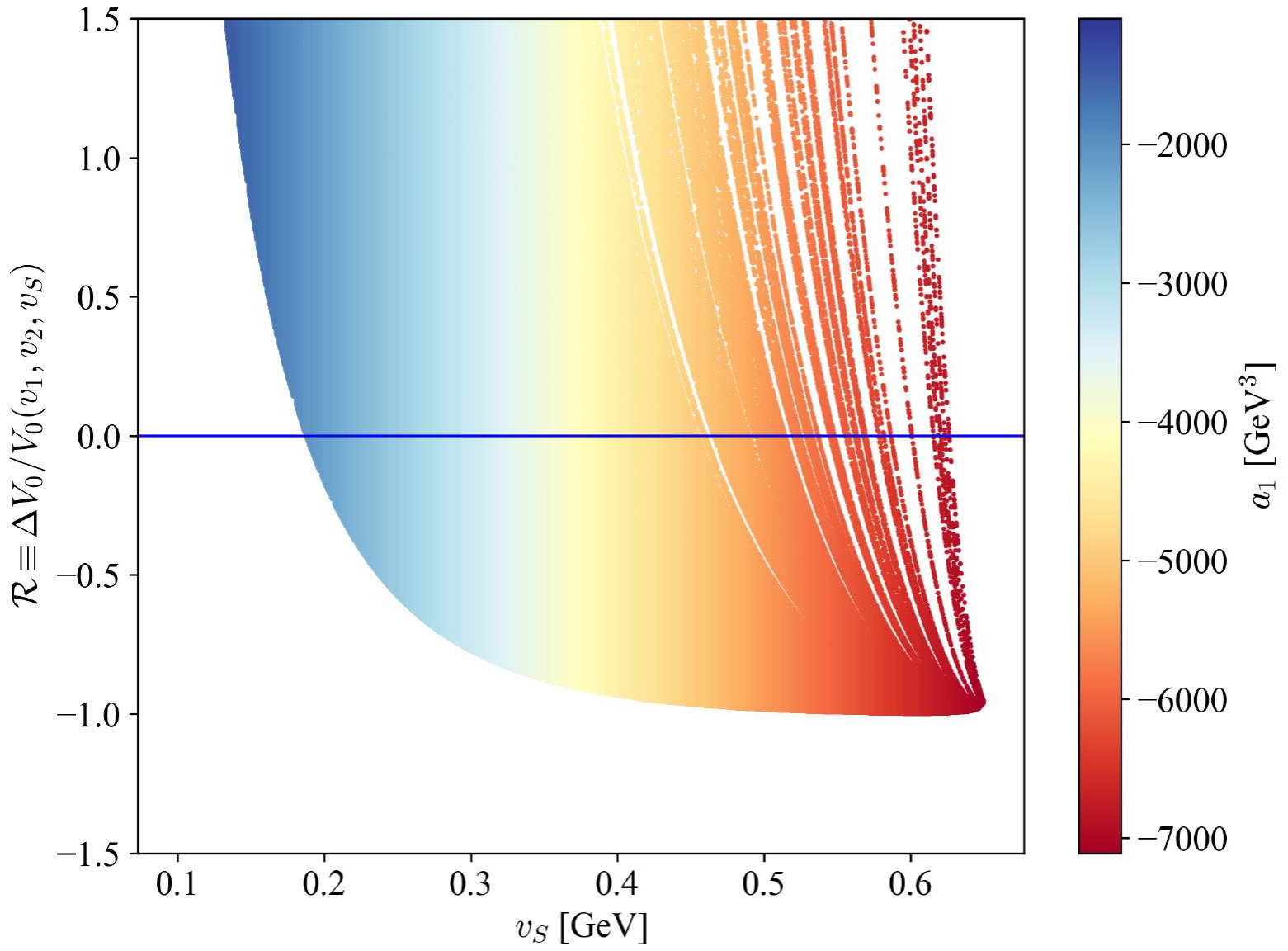}
\caption{Numerical results for $\mathcal{R} \equiv \Delta V_0 / V_0(v_1, v_2, v_S)$ as a function of the singlet VEV $v_S$, where the color indicates the value of the soft-breaking parameter $a_1$.
The left panel corresponds to $0<v_S<1000$ GeV, while the right panel focuses on the region with $v_S \lesssim 0.7~\mathrm{GeV}$.}
\label{fig:vS_scan}
\end{figure}
%----------------------------------------------------------------------------------------------------------------------------------

Fig.~\ref{fig:vS_scan} shows the numerical results for $\mathcal{R} \equiv \Delta V_0 / V_0(v_1, v_2, v_S)$ as a function of the singlet VEV $v_S$, with the color indicating the soft breaking parameter $a_1$.
In the left panel, a few points approach and slightly cross the $\mathcal{R} = 0$ line only in the region with very small $v_S$,
while no such points appear for larger $v_S$.
The right panel enlarges this small-$v_S$ region, making the behavior near $\mathcal{R} = 0$ more visible.
Note that the range of the parameter $a_1$ differs between the left and right panels.
As explained earlier, realizing the tree-level MPP condition requires large mixing parameters $\delta_{1,2}$, which are enhanced for smaller $v_S$.
Consequently, solutions close to $\mathcal{R} = 0$ appear exclusively at small $v_S$, as clearly seen in the figure\footnote{The parameter $a_1$ is taken to be of order the electroweak scale cubed, i.e., in the range $-(100)^3$--$(100)^3\,\mathrm{GeV}^3$. We have checked that extending this range does not qualitatively change the behavior of the results, and the points satisfying $\mathcal{R} = 0$ still appear only in the small-$v_S$ region.}.

\begin{comment}
We also comment on the behavior of the color distribution in the small-$v_S$ region of
Fig.~\ref{fig:vS_scan}.
$d_2$ is dominated by the term proportional to $a_1 / v_S^3$, as seen from Eq.~\eqref{d2}.
Since the theoretical constraints require $0 < d_2 < 4\pi$,  a small value of $v_S$ forces $a_1$ to fall within a very narrow range.
Therefore, in the small-$v_S$ region, $a_1$ is almost uniquely determined by the value of $v_S$, which is why the color distribution in the right panel appears well aligned for small $v_S$.
\end{comment}
We also comment on the behavior of the color distribution in the small-$v_S$ region of Fig.~\ref{fig:vS_scan}. 
$d_2$ is dominated by the term proportional to $a_1 / v_S^3$, as seen from Eq.~\eqref{d2}. 
Since the theoretical constraints require $0 < d_2 < 4\pi$, a small value of $v_S$ forces $a_1$ to fall within a very narrow range. 
Therefore, in the small-$v_S$ region, $a_1$ is almost uniquely determined by the value of $v_S$.
This is why the color distribution in the right panel appears well aligned for small $v_S$.

%----------------------------------------------------------------------------------------------------------------------------------
\begin{figure}[htpb]
\centering
\includegraphics[width=8cm]{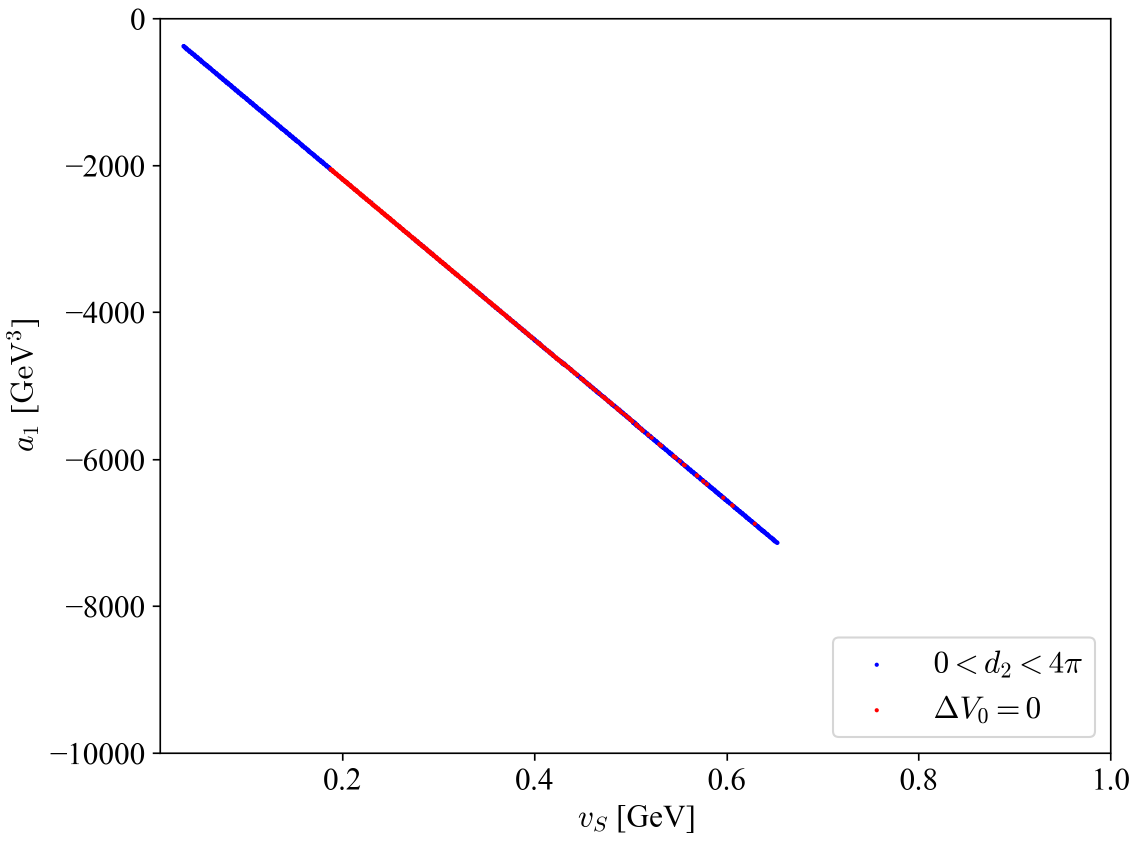}
\caption{
Parameter points in the $(v_S, a_1)$ plane.
Blue points satisfy the theoretical constraint $0<d_2<4\pi$, while red points additionally
fulfill the tree-level MPP condition $\Delta V_0=0$.}
\label{fig:vS_a1_line}
\end{figure}
%----------------------------------------------------------------------------------------------------------------------------------

To further illustrate this behavior, Fig.~\ref{fig:vS_a1_line} shows the parameter points in the $(v_S, a_1)$ plane.
The blue points represent the region satisfying the theoretical constraint $0 < d_2 < 4\pi$, while the red points indicate the subset of these points that
also fulfill the tree-level MPP condition $\Delta V_0 = 0$.
For small values of $v_S$, the allowed points align approximately along a straight line, reflecting the fact that $d_2$ is dominated by the term proportional to $a_1 / v_S^3$. As $v_S$ increases, however, the MPP condition itself ceases to be satisfied, even though the theoretical bound $0 < d_2 < 4\pi$ can still be fulfilled.
This shows that the MPP requirement provides a stronger restriction on the parameter space than the perturbativity bound on $d_2$, effectively setting an upper limit on $v_S$ in the degenerate scalar scenario.

Before closing this section, we comment on the implications for the EWPT.
A strong first-order EWPT, which is required for successful electroweak
baryogenesis (EWBG), is commonly characterized by the condition
\cite{Arnold:1987mh,Bochkarev:1987wf,Funakubo:2009eg}
\begin{align}
\frac{v_C}{T_C} \gtrsim 1 ,
\label{dcpcond}
\end{align}
where $T_C$ denotes the critical temperature at which the finite-temperature
effective potential develops two degenerate minima, and $v_C$ is the Higgs
vacuum expectation value at $T = T_C$.

In the 2HDMS, a first-order EWPT can, in principle, originate from two distinct
sources:
(i) the structure of the tree-level scalar potential, and
(ii) thermal loop effects that generate cubic terms in the finite-temperature
effective potential.
However, once the tree-level MPP condition is imposed, the electroweak and
singlet vacua become exactly degenerate at zero temperature.
As a consequence, the tree-level potential does not develop an energy barrier
between the two phases, and a tree-level-induced first-order EWPT is not
possible in the MPP setup.

In this situation, a first-order EWPT can arise only from thermal loop effects.
In the full one-loop finite-temperature effective potential, bosonic thermal
contributions generate cubic terms that can induce an energy barrier between
the symmetric and broken phases, potentially leading to a strong first-order
EWPT.
For completeness, the finite-temperature effective potential employed in our
analysis is given by
\begin{align}
V_{\mathrm{eff}}(\varphi_1, \varphi_2 , \varphi_S ; T)
&=
V_{0}(\varphi_1, \varphi_2 , \varphi_S)
+ \sum_i n_i
\left[
V_{\mathrm{CW}}(\bar{m}_i^2)
+ \frac{T^4}{2\pi^2}
I_{B,F}\!\left(\frac{\bar{m}_i^2}{T^2}\right)
\right],
\label{eff}
\end{align}
where $n_i$ is the number of degrees of freedom of particle $i$,
$V_{\mathrm{CW}}$ denotes the Coleman--Weinberg potential, and
$I_{B,F}$ are the thermal functions for bosons and fermions, respectively.

Finally, we note an important point regarding parameter dependence.
The tree-level MPP condition favors large values of the mixing parameters $\delta_{1,2}$,
since they enhance the separation between $v_S$ and $v_S'$ and help realize $\Delta V_0=0$.
In contrast, the degenerate scalar scenario discussed in Sec.~\ref{subsec:DSS} requires
$\delta_{1,2}$ to be sufficiently small in order to suppress the DM-nucleon scattering
amplitude.
These two requirements therefore act in opposite directions, making their compatibility
highly nontrivial. In the next section, we investigate this interplay by exploring concrete
benchmark points.

%%%%%%%%%%%%%%%%%%%%%%%%%%%%%%%%%%%%%%%%%%%%%%%%
%				DM phenomenology
%%%%%%%%%%%%%%%%%%%%%%%%%%%%%%%%%%%%%%%%%%%%%%%%

\section{DM phenomenology}\label{sec:DM}

We now turn to the DM phenomenology of the 2HDMS.
In this section, we investigate whether the parameter points that satisfy the tree-level MPP condition, as studied in Sec.~\ref{sec:MPP}, can also accommodate the results of the DM experiments. We select several representative benchmark points that satisfy the MPP condition and examine their predictions for the DM relic abundance and the spin-independent DM-nucleon scattering cross section.

%----------------------------------------------------------------------------------------------------------------------------------
\begin{table}[t]
\centering
\begin{tabular}{|c|c|c|c|c|c|c||c|c|}
\hline
& \multicolumn{6}{c||}{\textbf{Inputs}} & \multicolumn{2}{c|}{\textbf{Outputs}}
\\
\hline
BP & $m_{H_1}$ [GeV] & $m_{H_2}$ [GeV] & $m_{H_3}$ [GeV] & $m_{\chi}$ [GeV]
& $v_S$ [GeV]
& $a_1$ [GeV$^3$]
& $\delta_1$ & $\delta_2$
\\ \hline
BP1
& 125.0 & 124.5 & 124.0 & 62.5
& 0.20
& $-2194.76$
& 8.49 & 2.04
\\
BP2
& 125.0 & 124.5 & 124.0  & 62.5
& 0.40
& $-4411.71$
& 4.22 & 1.01
\\
BP3
& 125.0 & 124.5 & 124.0  & 62.5
& 0.60
& $-6607.28$
& 2.81 & 0.67
\\ \hline
\end{tabular}
\caption{Benchmark points satisfying the tree-level MPP condition. The three Higgs boson masses to be nearly degenerate within 0.5~GeV. All other input parameters not shown in this table are fixed to the same values as those given in Table~\ref{tab:input}.}
\label{tab:BPver1}
\end{table}
%----------------------------------------------------------------------------------------------------------------------------------

The benchmark points are summarized in Table~\ref{tab:BPver1}.
First, we fix the three Higgs boson masses to be nearly degenerate, with a 0.5~GeV separation, consistent with the degenerate scalar scenario.
Next, motivated by the results in Fig.~\ref{fig:vS_scan}, we select three representative values of the singlet VEV $v_S$ in the small-$v_S$ region favored by the MPP condition.
As seen in Fig.~\ref{fig:vS_a1_line}, imposing the tree-level MPP fixes the soft-breaking parameter $a_1$ almost uniquely once $v_S$ is chosen.
For each selected $v_S$, we therefore determine the corresponding value of $a_1$ that satisfies $\Delta V_0 = 0$, and compute the associated parameters $\delta_{1,2}$, which are crucial for the DM phenomenology and the MPP analysis.
According to Eqs.~\eqref{del1} and \eqref{del2}, the mixing parameters $\delta_{1,2}$ scale as $1/v_S$; therefore, increasing $v_S$ leads to a suppression of $\delta_{1,2}$.

The observed DM relic abundance reported by Planck~\cite{Planck:2018vyg} is
\begin{align}
\Omega_{\mathrm{DM}} h^2 = 0.1200 \pm 0.0012,
\label{relic}
\end{align}
which provides an upper bound on the relic density of the singlet-like DM particle $\chi$.
For our numerical analysis, we employ the public code \texttt{micrOMEGAs}~\cite{Belanger:2020gnr}, which computes both the relic abundance of $\chi$, $\Omega_\chi h^2$, and the spin-independent cross section $\sigma_{\mathrm{SI}}$.
As we will show below, in the parameter region favored by the MPP and the degenerate scalar scenario, the particle $\chi$ typically constitutes only a fraction of the total DM abundance, $\Omega_\chi < \Omega_{\mathrm{DM}}$.
In this case, the effective scattering rate in direct detection experiments must be rescaled by the fraction of $\chi$ in the Universe:
\begin{align}
\tilde{\sigma}_{\mathrm{SI}}
= \left( \frac{\Omega_\chi}{\Omega_{\mathrm{DM}}} \right)
\sigma_{\mathrm{SI}}.
\label{scale}
\end{align}
This rescaled cross section is then compared with the current upper limits from the LZ experiment~\cite{LZ:2024zvo}.
In computing the DM relic density and the DM--quark scattering cross section, we treat the DM mass $m_\chi$ as a free variable.

%----------------------------------------------------------------------------------------------------------------------------------
\begin{figure}[htpb]
\centering
\includegraphics[width=8cm]{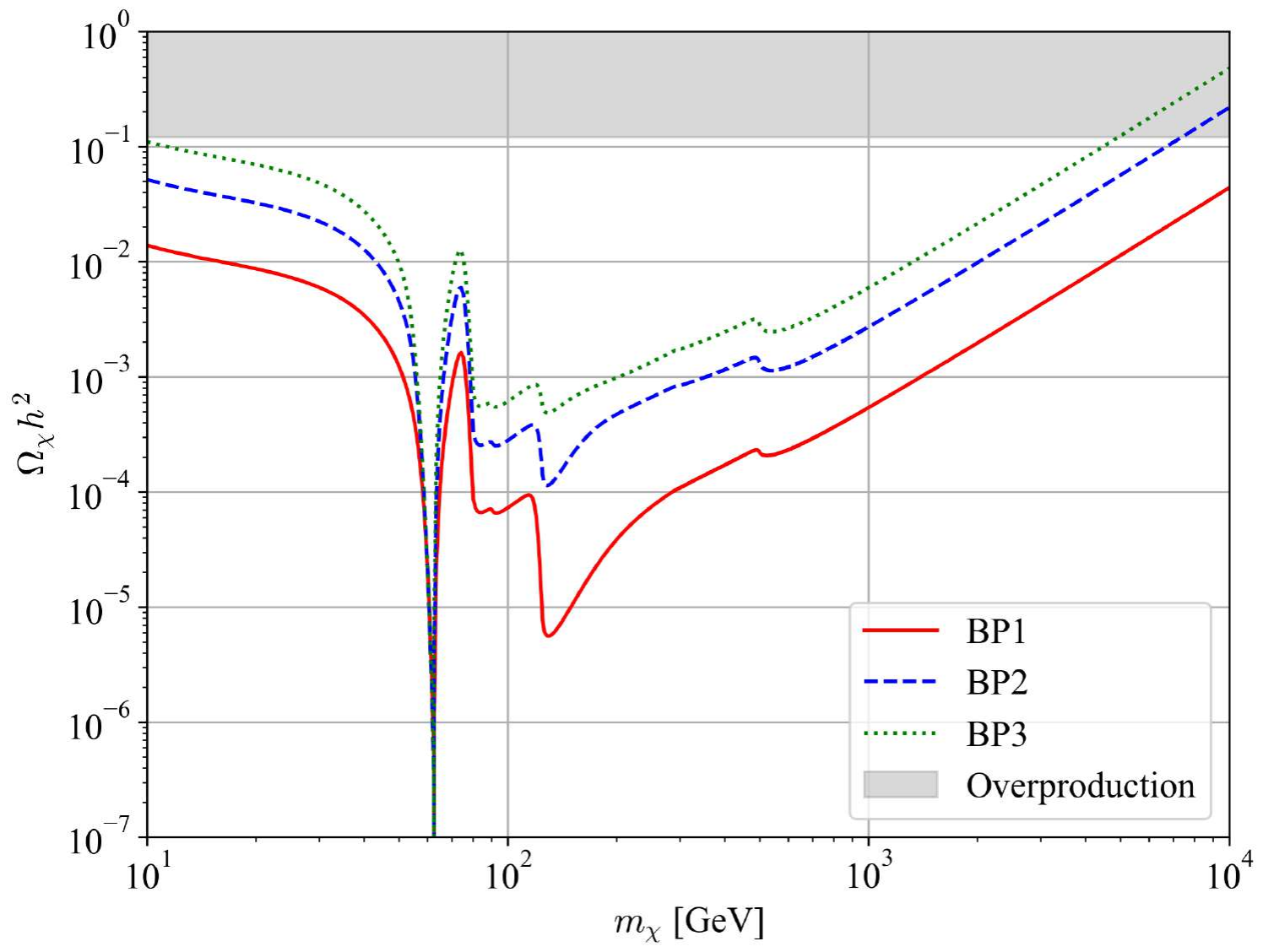}
\includegraphics[width=8cm]{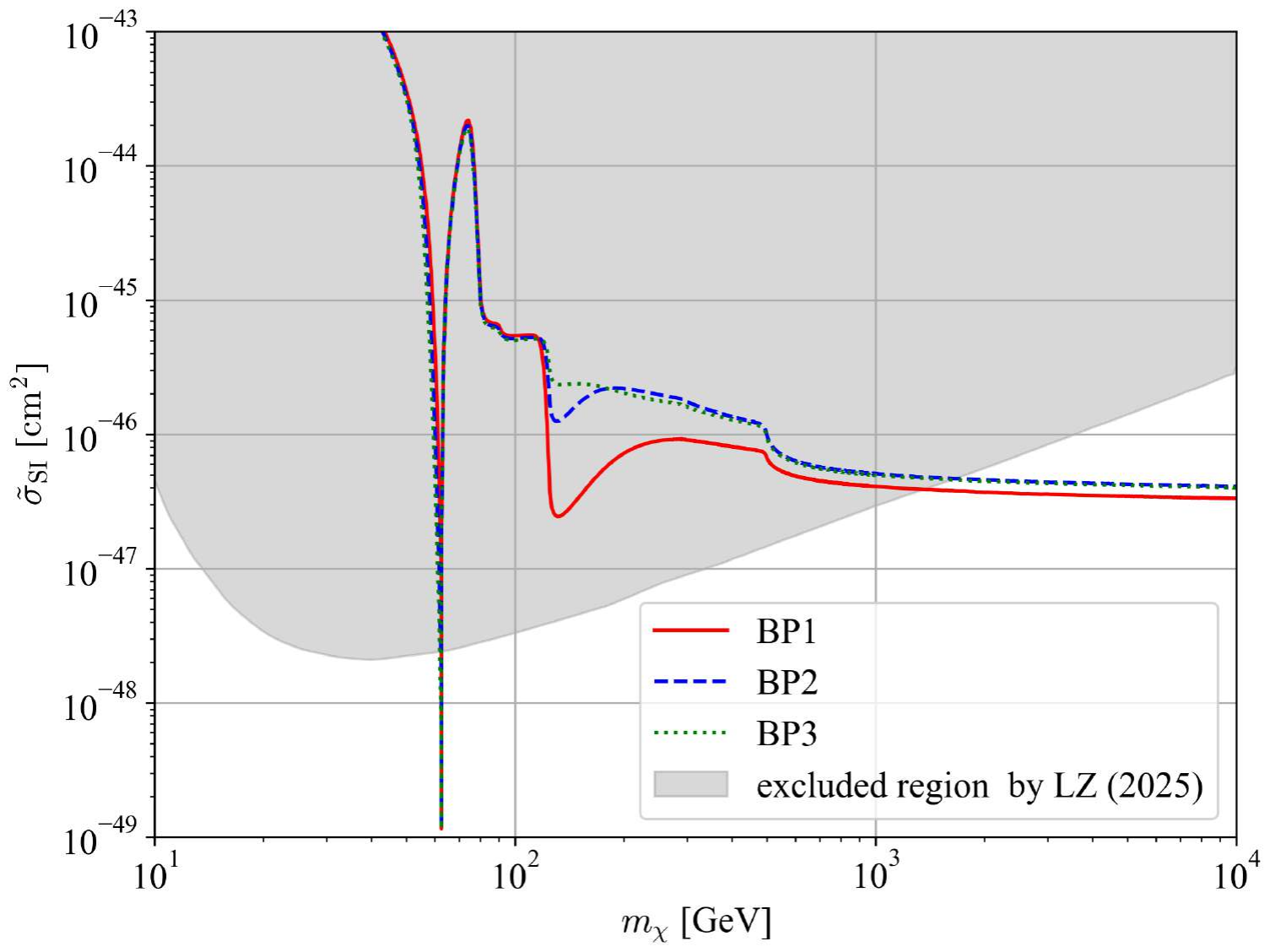}
\caption{
Left: Relic abundance $\Omega_\chi h^2$ for BP1 (red solid), BP2 (blue dashed), and BP3 (green dotted).
The gray-shaded region corresponds to DM overproduction.
Right: Scaled spin-independent direct-detection cross section
$\tilde{\sigma}_{\mathrm{SI}}$ (see Eq.~\eqref{scale}) for the same benchmark points.
The gray shaded region is excluded by the LZ experiment~\cite{LZ:2024zvo}.
}
\label{DM_BP123}
\end{figure}
%----------------------------------------------------------------------------------------------------------------------------------

Fig.~\ref{DM_BP123} shows the relic abundance (left) and the scaled spin-independent DM--nucleon scattering cross section (right) as functions of the DM mass~$m_\chi$.
As seen in the left panel, the size of the relic abundance is strongly correlated with the mixing parameters~$\delta_{1,2}$ (See Table~\ref{tab:BPver1}).
For BP1, where $\delta_{1,2}$ takes relatively large values, the DM annihilation cross section is enhanced, and the relic abundance is significantly reduced compared with BP2 and BP3.
The sharp dip around $m_\chi \simeq m_{H_i}/2 \simeq 62.5~\mathrm{GeV}$ corresponds to the $s$-channel Higgs-resonant annihilation.

The right panel of Fig.~\ref{DM_BP123} shows the scaled scattering cross section $\tilde{\sigma}_{\mathrm{SI}}$ and therefore reflects the shape of the relic density curve.
At first sight, it may seem surprising that the direct-detection rate is not strongly suppressed, despite the use of the degenerate scalar scenario discussed in Sec.~\ref{subsec:DSS}.
The reason is that the suppression mechanism relies on small values of $\delta_{1,2}$, whereas the tree-level MPP condition favors larger $\delta_{1,2}$ in order to enhance the separation between $v_S$ and $v_S'$ and realize $\Delta V_0 = 0$.
Thus, the requirements for $\delta_{1,2}$ from the degenerate scalar scenario and from the MPP act in opposite directions.

%----------------------------------------------------------------------------------------------------------------------------------
\begin{table}[t]
\centering
\begin{tabular}{|c|c|c|c|c|c|c||c|c|}
\hline
& \multicolumn{6}{c||}{\textbf{Inputs}} & \multicolumn{2}{c|}{\textbf{Outputs}}
\\
\hline
BP & $m_{H_1}$ [GeV] & $m_{H_2}$ [GeV] & $m_{H_3}$ [GeV] & $m_{\chi}$ [GeV]
& $v_S$ [GeV]
& $a_1$ [GeV$^3$]
& $\delta_1$ & $\delta_2$
\\ \hline
BP4
& 125.0 & 124.5 & 124.0 & 62.5
& 0.63
& $-6867.62$
& 2.71 & 0.65
\\
BP5
& 125.0 & 124.75 & 124.5 & 62.5
& 0.31
& $-3388.83$
& 2.77 & 0.66
\\
BP6
& 125.0 & 124.9 & 124.8 & 62.5
& 0.12
& $-1376.29$
& 2.74 & 0.66
\\ \hline
\end{tabular}
\caption{
Benchmark points with increasingly degenerate neutral scalar masses.
The mass differences among $H_1$, $H_2$, and $H_3$ are 0.50~GeV (BP4), 0.25~GeV (BP5), and 0.10~GeV (BP6), respectively.
All other input parameters not shown in this table are fixed to the same values as those given in Table~\ref{tab:input}.
}
\label{tab:BPver2}
\end{table}
%----------------------------------------------------------------------------------------------------------------------------------

In the previous subsection, we examined benchmark points in which the three neutral Higgs
bosons were nearly degenerate with mass differences of about $0.5~\mathrm{GeV}$.
These points satisfy the tree-level MPP condition, but do not fully realize
the suppression mechanism of the degenerate scalar scenario, since the MPP prefers
relatively large values of $\delta_{1,2}$.

A natural question is therefore whether a stronger mass degeneracy among the Higgs bosons could improve the situation.
To address this, we introduce additional benchmark points, BP4, BP5, and BP6, shown in Table~\ref{tab:BPver2}.
Here, the Higgs masses are chosen to be more tightly clustered, with differences
reduced to 0.50~$\mathrm{GeV}$ (BP4), 0.25~$\mathrm{GeV}$ (BP5), and 0.10~$\mathrm{GeV}$ (BP6).
As can be seen from Table~\ref{tab:BPver2}, the resulting values of the mixing parameters $\delta_{1,2}$ are almost identical for BP4, BP5, and BP6.
This is because, for each choice of the Higgs mass difference, the singlet VEV $v_S$ is chosen to be as large as possible while still satisfying the tree-level MPP condition.
Such a choice effectively corresponds to selecting the minimal values of $\delta_{1,2}$ compatible with the MPP.
Indeed, from Eqs.~\eqref{del1} and \eqref{del2}, once $\delta_{1,2}$ are fixed, the singlet VEV $v_S$ is uniquely determined for a given pattern of Higgs mass differences. As a result, increasing the degree of mass degeneracy does not provide additional freedom to reduce $\delta_{1,2}$ when the MPP condition is imposed.

%----------------------------------------------------------------------------------------------------------------------------------
\begin{figure}[htpb]
\centering
\includegraphics[width=8cm]{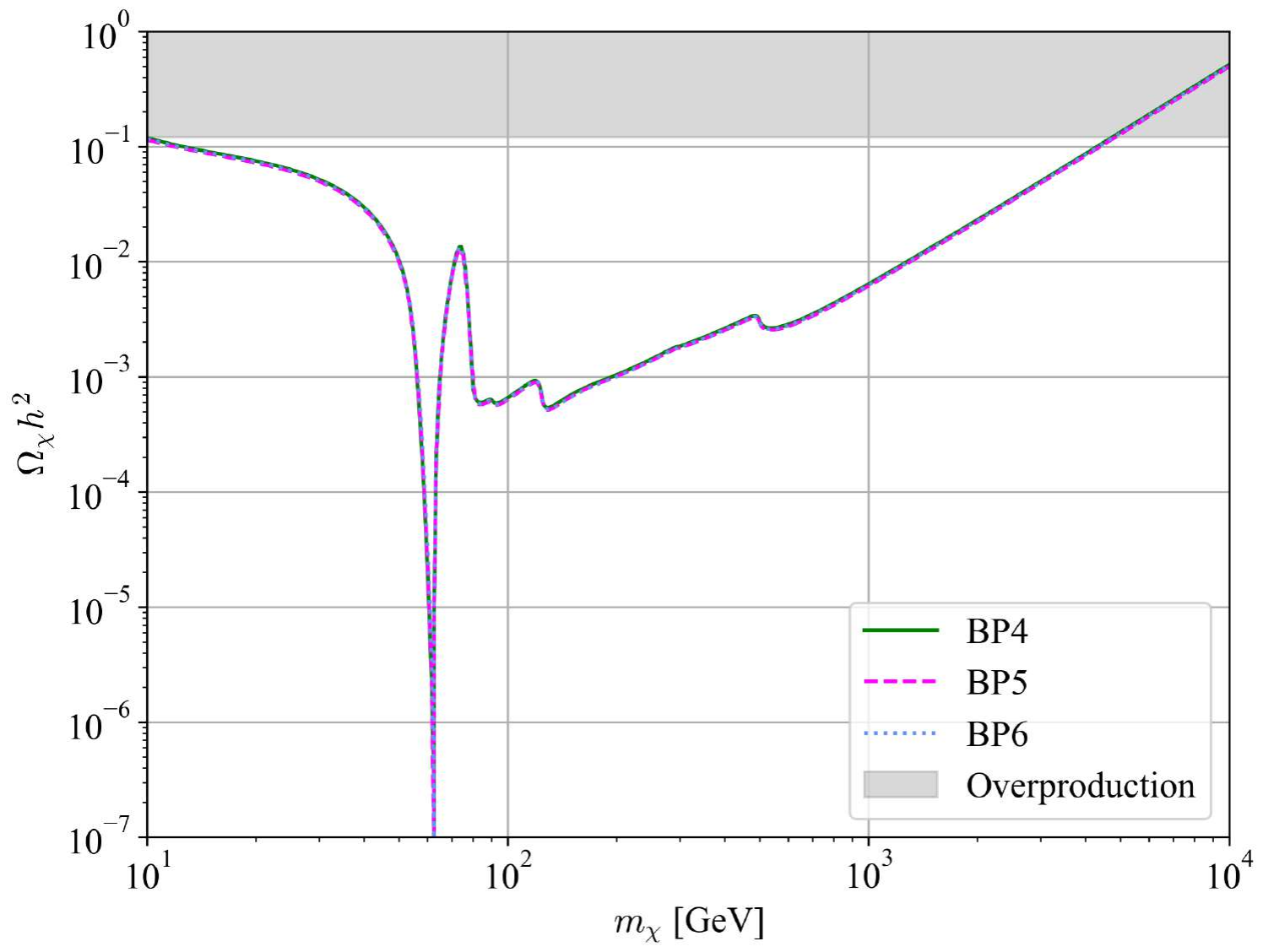}
\includegraphics[width=8cm]{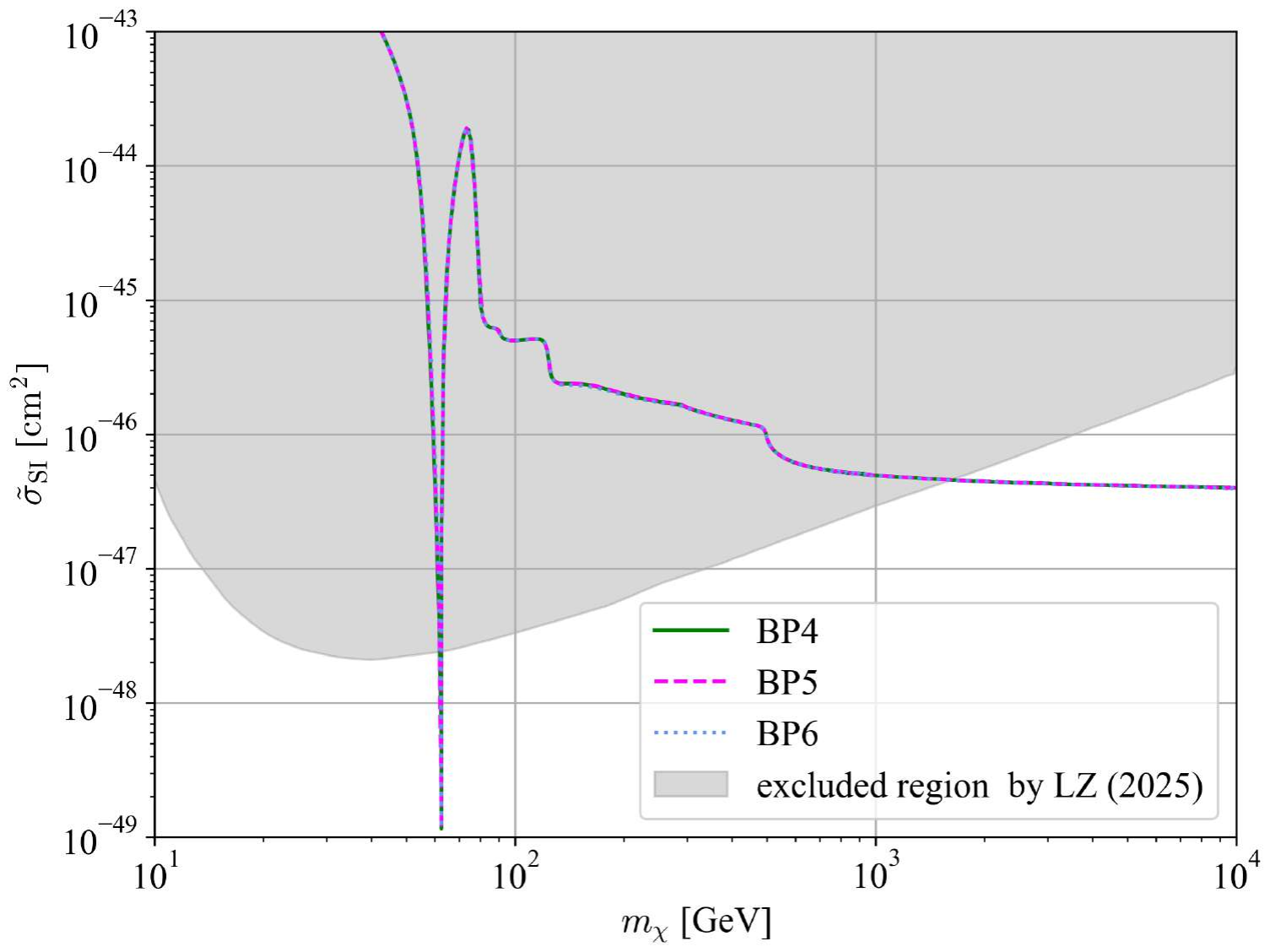}
\caption{
Left: Relic abundance $\Omega_\chi h^2$ for BP4 (green solid), BP5 (magenta dashed), and BP6 (light blue dotted).
The gray-shaded region corresponds to DM overproduction.
Right: Scaled spin-independent direct-detection cross section
$\tilde{\sigma}_{\mathrm{SI}}$ (see Eq.~\eqref{scale}) for the same benchmark points.
The gray shaded region is excluded by the LZ experiment~\cite{LZ:2024zvo}.
}
\label{DM_BP456}
\end{figure}
%----------------------------------------------------------------------------------------------------------------------------------

The relic abundance and direct detection predictions for BP4, BP5, and BP6 are shown in Fig.~\ref{DM_BP456}.
As discussed above, since the mixing parameters $\delta_{1,2}$ take almost identical values for BP4, BP5, and BP6, the corresponding curves in Fig.~\ref{DM_BP456} nearly overlap and are hardly distinguishable.
As a result, increasing the level of mass degeneracy does not lead to additional suppression of the DM scattering amplitude.

It is worth emphasizing that viable parameter space remain.
Two regions allow simultaneous satisfaction of the tree-level MPP condition,
the DM relic abundance requirement, and the DM direct detection constraint:
(i) the resonance region near $m_\chi \simeq 62.5~\mathrm{GeV}$, and
(ii) the heavy mass regime with $m_\chi = \mathcal{O}(1\text{--}10)\,\mathrm{TeV}$.
These regions provide concrete examples in which the MPP and DM phenomenology
can be compatible within the 2HDMS.

\begin{table}[t]
\centering
\begin{tabular}{|c|c|c|c|c|c|c|}
\hline
 & BP1 & BP2 & BP3 & BP4 & BP5 & BP6 \\ \hline
$v_C/T_C$
& $\frac{237.9}{69.7}=3.4$
& $\frac{237.9}{69.7}=3.4$
& $\frac{237.9}{69.7}=3.4$
& $\frac{238.2}{68.4}=3.5$
& $\frac{238.1}{69.1}=3.4$
& $\frac{238.0}{69.5}=3.4$ \\ \hline
\end{tabular}\caption{Ratio $v_C/T_C$ at the critical temperature for BP1--BP6, evaluated using \texttt{cosmoTransitions}~\cite{Wainwright:2011kj}. The Parwani resummation method~\cite{Parwani:1991gq} is used in this analysis.}
\label{tab:TCvC}
\end{table}
%----------------------------------------------------------------------------------------------------------------------------------

Finally, we comment on the strength of the EWPT at the quantitative level.
The calculations are carried out using \texttt{CosmoTransitions}~\cite{Wainwright:2011kj}, based on the one-loop finite-temperature effective potential given in Eq.~\eqref{eff}.
At finite temperature, however, bosonic multi-loop contributions can spoil the validity of naive perturbation theory.
To properly account for these effects, thermal resummation is required.
In this work, we adopt the Parwani resummation scheme~\cite{Parwani:1991gq}, in which thermal corrections are incorporated by resumming all Matsubara frequency modes.
In a tree-level-driven first-order EWPT, the strength of the transition is highly sensitive to the mixing parameters $\delta_{1,2}$, which control the
structure of the tree-level potential.
However, in the present setup, the tree-level MPP condition is imposed, and therefore, a tree-level-induced first-order phase transition does not occur.

Instead, the first-order EWPT is generated by thermal loop effects at the one-loop level as discussed in Sec.~\ref{sec:MPP}.
In this case, the dominant contribution to the cubic term in the finite-temperature effective potential arises from bosonic thermal loops, and the strength of the
transition is mainly controlled by the bosonic mass spectrum.
Since our benchmark points are chosen in a region where the scalar masses are nearly degenerate, there is no significant variation in the relevant bosonic
masses among BP1--BP6.
As a result, the values of $v_C/T_C$ show only mild differences across the benchmark point as shown in Table~\ref{tab:TCvC}.
We find that all benchmark points satisfy the condition for a strong first-order EWPT in Eq.~\eqref{dcpcond}.

We also comment on the heavy DM region. As seen in Fig.~3, for large values of $m_\chi$, e.g., $m_\chi = 2000\,\mathrm{GeV}$, the DM relic abundance falls below the observed value, and the DM–nucleon scattering cross section remains consistent with current direct detection constraints.
As discussed above, the key parameters relevant for the MPP condition and the EWPT are the mixing parameters $\delta_{1,2}$. While $\delta_{1,2}$ are determined by Eqs.~\eqref{del1} and \eqref{del2}, the DM mass $m_\chi$ is controlled by the soft breaking parameters $a_1$ and $b_1$ as shown in Eq.~\eqref{mDM}. Therefore, $m_\chi$ has only a minor impact on $\delta_{1,2}$, and varying $m_\chi$ does not significantly affect the MPP condition or the structure of the EWPT.

%%%%%%%%%%%%%%%%%%%%%%%%%%%%%%%%%%%%%%%%%%%%%%%%
%				Summary
%%%%%%%%%%%%%%%%%%%%%%%%%%%%%%%%%%%%%%%%%%%%%%%%

\section{Summary}\label{sec:sum}

In this work, we have studied the 2HDMS, focusing on the implications of imposing the tree-level MPP.
The scalar potential of the model naturally possesses two distinct vacua along the electroweak and singlet directions, and we required these vacua to be degenerate at tree level.
This tree-level MPP condition fixes a nontrivial relation among the parameters.
We demonstrated that the MPP requirement favors large values of the mixing parameters $ \delta_{1,2} $, because they enhance the separation between $v_S$ and $v_S'$ and enable the realization of $\Delta V_0 = 0$.

The imaginary component of the singlet field serves as a stable WIMP DM candidate.
Its interactions with nucleons arise from $t$-channel exchange of the three neutral Higgs bosons $H_{1,2,3}$.
When the masses of these scalars are nearly degenerate, the orthogonality of the mixing matrix leads to cancellations among the scattering amplitudes, naturally suppressing the spin-independent DM-quark cross section.
This mechanism, known as the degenerate scalar scenario, offers a compelling way to satisfy stringent direct detection constraints.
However, the degenerate scalar scenario requires the mixing parameters $ \delta_{1,2} $ to be sufficiently small for the amplitude cancellation to be effective.
As a result, the MPP and the degenerate scalar scenario impose mutually competing requirements on $ \delta_{1,2} $, making their simultaneous realization highly nontrivial.

We explored this tension by constructing explicit benchmark points that satisfy the tree-level MPP condition.
The tree-level MPP requires large values of the mixing parameters $\delta_{1,2}$, and achieving such values in turn demands a small singlet VEV $v_S$.
As anticipated, this makes the suppression mechanism of the degenerate scalar scenario less effective.
To examine whether stronger mass degeneracy could improve the situation, we further considered cases in which the neutral Higgs bosons are even more tightly degenerate in mass.
However, the MPP effectively fixes $v_S$ for a given pattern of mass differences, thereby determining the size of $\delta_{1,2}$.
As a consequence, increasing the level of mass degeneracy does not lead to additional suppression of the DM direct-detection rate.
Nevertheless, we found that two parameter regions remain in which all constraints can be satisfied simultaneously:
(i) the Higgs-resonance region with $m_\chi \simeq 62.5~\mathrm{GeV}$, and
(ii) the heavy-mass regime with $m_\chi = \mathcal{O}(1\text{--}10)~\mathrm{TeV}$.

We also examined the implications of the tree-level MPP for the EWPT.
Although the MPP forbids a tree-level-driven first-order transition due to the exact vacuum degeneracy at zero temperature, we found that thermal loop effects
at the one-loop level can still generate a strong first-order EWPT. Indeed, all benchmark points considered in this work satisfy the conventional criterion for a strong first-order EWPT.

To summarize, imposing the tree-level MPP strongly constrains the structure of the scalar potential, most notably by requiring the mixing parameters $\delta_{1,2}$ to be large.
In contrast, the degenerate scalar scenario introduced to suppress the DM spin-independent scattering rate essentially requires the same parameters $\delta_{1,2}$ to be sufficiently small.
Thus, the tree-level MPP and the degenerate scalar scenario impose mutually conflicting demands on $\delta_{1,2}$, making their simultaneous realization highly nontrivial.
Nevertheless, in this work, we have shown that viable parameter regions remain, in which the results of the DM experiments are reproduced while a strong first-order EWPT can still be achieved.

%%%%%%%%%%%%%%%%%%%%%%%%%%%%%%%%%%%%%%%%%%%%%%%%
%				Acknowledgments
%%%%%%%%%%%%%%%%%%%%%%%%%%%%%%%%%%%%%%%%%%%%%%%%

\begin{acknowledgments}
The work of GCC is supported by JSPS KAKENHI Grant No. 22K03616. The work of CI is supported by the National Natural Science Foundation of China (NNSFC) Grant No.12475111.

\end{acknowledgments}

%%%%%%%%%%%%%%%%%%%%%%%%%%%%%%%%%%%%%%%%%%%%%%%%
%							Appendix
%%%%%%%%%%%%%%%%%%%%%%%%%%%%%%%%%%%%%%%%%%%%%%%%
\appendix

\section{Input and output parameters}\label{app:para}
The following shows the relationship between the input and output parameters described in Table~\ref{tab:inout} except for $\delta_1,\delta_2$ and $d_2$.

\begin{align}
m_1^2 &= \frac{m_3^2 v_2}{v_1} - \frac{\lambda_1 v_1^2}{2} - \frac{\lambda_{345} v_2^2}{2} -\frac{\delta_1 v_S^2}{4}, \\
m_2^2 &= \frac{m_3^2 v_1}{v_2} - \frac{\lambda_2 v_2^2}{2} - \frac{\lambda_{345} v_1^2}{2} -\frac{\delta_2 v_S^2}{4}, \\
\lambda_1 &= \frac{1}{v_1^2} \left(\sum_{i=1}^3 O_{1i}^2 m_{H_i}^2 - \frac{m_3^2 v_2}{v_1} \right),\\
\lambda_2 &= \frac{1}{v_2^2} \left(\sum_{i=1}^3 O_{2i}^2 m_{H_i}^2- \frac{m_3^2 v_1}{v_2} \right),\\
\lambda_3 &= \frac{1}{v_1 v_2} \left(\sum_{i=1}^3 O_{1i} O_{2i} m_{H_i}^2  + m_3^2 \right) - \lambda_{4} - \lambda_{5}, \\
\lambda_{4} &= \frac{2}{v^2} \left( \frac{m_3^2}{\sin\beta \cos\beta} - m_{H^\pm}^2 \right) - \lambda_{5}, \\
\lambda_{5} &= \frac{1}{v^2}\left(\frac{m_3^2}{\sin\beta \cos\beta} - m_A^2\right),\\
b_2 &= \frac{-4 \sqrt{2} a_1 - 2 b_1 v_S - \delta_1 v_1^2 v_S - \delta_2 v_2^2 v_S - d_2 v_S^3}{2 v_S}, \\
b_1 &= -m_{\chi}^2 - \frac{\sqrt{2} a_1}{v_S}.
\end{align}

%%%%%%%%%%%%%%%%%%%%%%%%%%%%%%%%%%%%%%%%%%%%%%%%
%					Reference
%%%%%%%%%%%%%%%%%%%%%%%%%%%%%%%%%%%%%%%%%%%%%%%%

\bibliography{biblist}

\end{document}